\documentclass[aps,epsfig,prb,unsortedaddress,twocolumn,footinbib]{revtex4}
\usepackage{amsmath}
\usepackage{graphicx}

\begin{document}

\title{Theory of intraband plasmons in doped carbon nanotubes:\\
rolled surface-plasmons of graphene}

\author{Ken-ichi Sasaki}
\email{sasaki.kenichi@lab.ntt.co.jp}
\affiliation{NTT Basic Research Laboratories, NTT Corporation,
3-1 Morinosato Wakamiya, Atsugi, Kanagawa 243-0198, Japan}

\author{Shuichi Murakami}
\affiliation{Department of Physics, Tokyo Institute of Technology,
2-12-1 Ookayama, Meguro, Tokyo 152-8551, Japan}

\author{Hideki Yamamoto}
\affiliation{NTT Basic Research Laboratories, NTT Corporation,
3-1 Morinosato Wakamiya, Atsugi, Kanagawa 243-0198, Japan}

\date{\today}
 
\begin{abstract}
 A single-wall carbon nanotube possesses
 two different types of plasmons specified by the wavenumbers in the
 azimuthal and axial directions.
 The azimuthal plasmon that is caused by interband transitions has
 been studied, while the effect of charge doping is unknown.
 In this paper, we show that when nanotubes are heavily doped, intraband
 transitions cause the azimuthal plasmons to appear as a plasmon
 resonance in the near-infrared region of the absorption spectra,
 which is absent for light doping due to the screening effect.
 The axial plasmons that are inherent in the cylindrical waveguide
 structures of nanotubes, account for the absorption peak of the
 metallic nanotube observed in the terahertz region.
 The excitation of axial (azimuthal) plasmons requires a linearly
 polarized light parallel (perpendicular) to the tube's axis.
\end{abstract}

\pacs{78.67.Ch, 73.63.fg, 73.20.Mf}
\maketitle


A carbon nanotube (CNT) has a notable optical property.~\cite{iijima91,saito92apl}
Namely, a CNT exhibits the absorption peaks of 
light with linear polarization parallel to the tube's axis, 
but not for the linear polarization perpendicular to it.~\cite{ichida04}
This optical anisotropy of a CNT is essential for 
characterizing a sample using Raman spectroscopy and for realizing
polarized optical devices.~\cite{Dresselhaus2010}
In this paper we show that when a CNT is heavily doped, many of the
absorption peaks of parallel polarized light disappear due to the Pauli
exclusion principle, and this causes an absorption peak of a
perpendicularly polarized light to appear in the near-infrared region.
This peak is a plasmon resonance that 
corresponds to the surface plasmon of graphene with the wavelength
on the order of nanometer.~\cite{stern67,Wunsch2006,hwang07,Jablan2009} 

\begin{figure}[htbp]
 \begin{center}
  \includegraphics[scale=0.4]{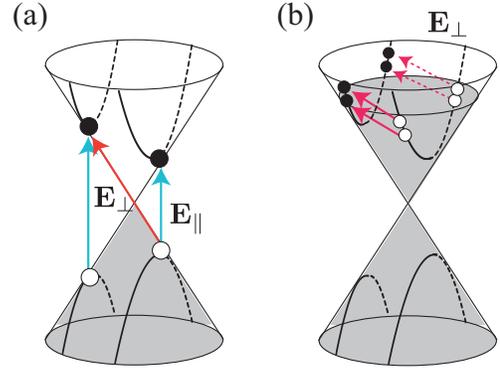}
 \end{center}
 \caption{(color online)
 (a) The optical selection rule allows interband transitions in an
 undoped CNT. Parallel (perpendicularly) polarized light ${\bf
 E}_\parallel$ (${\bf E}_\perp$) induces direct (indirect) interband
 transitions. However, the indirect interband transition is suppressed
 by the screening effect.
 (b) Collective excitations that are composed of intraband
 electron-hole pairs in a heavily doped CNT cause absorption of ${\bf
 E}_\perp$.
 }
 \label{fig:optical}
\end{figure}

The light polarization dependence of an undoped CNT is explained in terms
of the optical selection rule and screening effect.~\cite{Ajiki1994}
The selection rule states that a parallel (perpendicularly)
polarized light induces direct (indirect) interband transitions, see
Fig.~\ref{fig:optical}(a).~\cite{Sasaki2011}
Although indirect transition of perpendicularly polarized light is
allowed by the selection rule, it is suppressed by the screening effect
by a plasma mode whose frequency $\omega_p$ is much larger than that of 
the excitation of an indirect transition $\omega$ (i.e., $\omega_p \gg \omega$).
Namely, the external electric field (${\bf E}_\perp^{\rm ext}$) is
screened to give a small total electric field (${\bf E}_\perp$):
\begin{align}
 {\bf E}_\perp(\omega) = \frac{ {\bf E}_\perp^{\rm ext}(\omega) }{
 1-\left(\frac{\omega_p}{\omega} \right)^2 }.
 \label{eq:scEfield}
\end{align}
By assuming that $\omega_{p}$ of 
is approximately 5 eV for $\pi$-plasmon in an undoped CNT,~\cite{Stephan2002,GarciadeAbajo2010}
${\bf E}_\perp(\omega)$ is strongly suppressed and there is almost no
absorption of near-infrared light.

\begin{figure}[htbp]
 \begin{center}
  \includegraphics[scale=0.3]{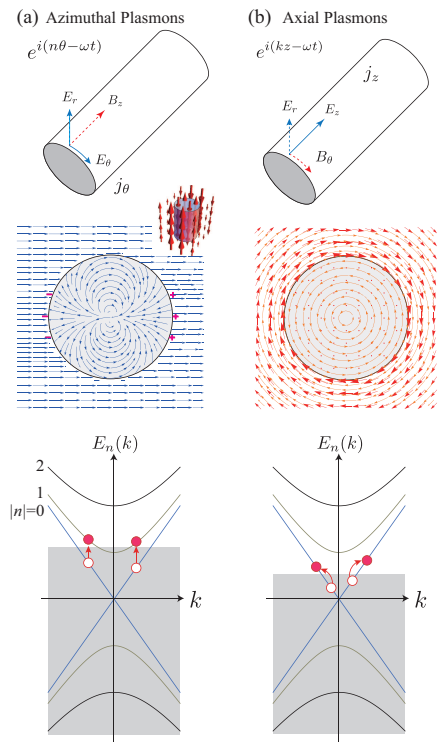}
 \end{center}
 \caption{(color online)
 Plasmon modes of a single-wall CNT.
 (a,top) 
 The azimuthal TM mode is characterized by the three nonzero components
 of the EM fields $(E_r,E_\theta,B_z)$, 
 where the electric (magnetic) field is denoted by a blue (red) arrow.
 A solid (dashed) arrow signifies that the vector is continuous
 (discontinuous) at the surface of a CNT.~\cite{Sasaki2016}
 (a,middle)
 A streamline plot of the electric field for the fundamental mode
 ($n=1$) is shown at the cross-section of a CNT.
 ``$+/-$'' represent positive/negative charges.
 The TM mode does not generate a nonzero total magnetic flux and
 the Aharonov-Bohm effect does not appear.
 (a,bottom) Some microscopic processes (electron-hole pair creation) that
 constitute the current $j_\theta$ are illustrated for a metallic CNT.
 The region below the Fermi energy is shown in gray and the energy
 bands with the same band index $n$ are denoted by the same color.
 (b) The nonzero EM components (top), the vector of the magnetic field
 shown in red and its linestream plot shown in orange (middle), and the
 processes (bottom) of the axial TM mode.
 }
 \label{fig:TMTE-cnt}
\end{figure}

The heavy doping into a CNT changes the situation:
doping can activate intraband transitions between two energy sub-bands
(with different orbital angular momentums around the tubule axis)
by a weak perturbation at low energy (see Fig.~\ref{fig:optical}(b)).
We find that the azimuthal motion of the electrons due to these intraband
transitions causes a plasma mode to appear in a heavily doped
CNT. 
This plasma mode originates from the surface plasmons in
graphene studied in Ref.~\onlinecite{stern67,Wunsch2006,hwang07,Jablan2009}.
The angular frequency of the surface plasmons in graphene is given by
\begin{align}
 \omega^{sp}_{k} = \sqrt{\frac{\sigma |{\bf k}|}{2\epsilon \tau}}-\frac{i}{2\tau},
 \label{eq:omegasp}
\end{align}
where $\epsilon$ is the permittivity of the surrounding material,
$\tau$ is the relaxation time, $\sigma$ is the static conductivity, 
and ${\bf k}$ is a two-dimensional
wavevector.~\cite{stern67,Wunsch2006,hwang07,Jablan2009}
Considering a CNT instead of graphene, we see that
there are two distinct plasmon modes, 
depending on the orientations of the wavevector ${\bf k}$;
one mode with ${\bf k}$ in the azimuthal direction of a cylinder
is called azimuthal plasmon mode, and the other with ${\bf k}$ parallel
to the axial direction is an axial plasmon mode
(see the top panels of Fig.~\ref{fig:TMTE-cnt}).
In the cylindrical coordinates $(r,\theta,z)$, 
they have the form of the plane waves $e^{i(n\theta-\omega t)}$ and
$e^{i(k z-\omega t)}$, respectively.
The azimuthal mode shown in Fig.~\ref{fig:TMTE-cnt}(a) has only three
components of the electromagnetic (EM) fields ($E_r,E_\theta,B_z$).
The unique nonzero magnetic field $B_z$ is perpendicular to the
azimuthal direction.
Similarly, the axial mode shown in Fig.~\ref{fig:TMTE-cnt}(b) has a unique
nonzero magnetic field $B_\theta$ that is perpendicular to the
axial direction.
These two modes are transverse magnetic (TM) modes.

The theoretical results presented in this paper were obtained
analytically by solving Maxwell equations with Ohm's law, i.e.
$j_{\theta}= \sigma_{\theta}(\omega) E_{\theta}$ for the azimuthal mode,
and $j_{z}= \sigma_{z}(\omega) E_{z}$ for the axial
mode.~\cite{nakayama74}
Let us
mention here that the Drude model is used to specify the dynamical
conductivity $\sigma_{\theta,z}(\omega)$ as
$\sigma_{\theta,z}(\omega)=\sigma_{\theta,z}/(1-i\omega \tau_{\theta,z})$, where the frequency 
dependence is controlled by the electronic relaxation times
$\tau_{\theta,z}$,
and $\sigma_{\theta,z}(\omega)$ reduces to the static conductivities
$\sigma_{\theta,z}$ in the zero-frequency limit. 
The application of Drude model to a heavily doped CNT is partly
justified by the polarization function.~\footnote{See supplemental material at 
[URL will be inserted by AIP] for details.}

We now describe properties of azimuthal plasmons shown in Fig.~\ref{fig:TMTE-cnt}(a).
The calculated angular frequencies are 
\begin{align}
 \omega_{n}^{\rm TM} = \sqrt{\frac{\sigma_\theta |n|}{2\epsilon \tau_\theta r_0}}
 - \frac{i}{2\tau_\theta}, 
 \label{eq:omega_TM}
\end{align}
where $n/r_0$ is the wavenumber ($n$ is a nonzero
integer and $r_0$ is the radius of the CNT).
Here, for the sake of clarity,
the same materials with the permittivity $\epsilon$
are assumed to occupy the inner and outer spaces of the CNT.
When different
materials occupy the inner and outer spaces of the CNT,
Eq.~(\ref{eq:omega_TM}) can be generalized by replacing $\epsilon$ with
$(\epsilon_{\rm out} + \epsilon_{\rm in})/2$, 
where $\epsilon_{\rm out}$ ($\epsilon_{\rm in}$) is the permittivity of
the outer (inner) space.
We note that Eq.~(\ref{eq:omega_TM}) is identical to Eq.~(\ref{eq:omegasp}),
if $n/r_0$ is replaced with $|{\bf k}|$.

Despite the similarity between 
the dispersion relations of a CNT and graphene,
their EM fields exhibit different asymptotic behavior.
Namely, the amplitude of the azimuthal TM mode
in a CNT behaves 
as $\propto 1/\sqrt{r}$ (extended)
and this mode is not a localized surface plasmon,
while the surface plasmon of graphene is exponentially
localized in the direction normal to the layer.~\cite{nakayama74}
The configuration of the electric field near a CNT 
is illustrated in Fig.~\ref{fig:TMTE-cnt}(a,middle).
We choose the fundamental mode with $n=1$ for the plot,
because it is the principal state excited by a perpendicular
polarization light.~\footnote{The excitation of the
fundamental mode by perpendicular polarization matches the selection
rule for the optical transitions in CNTs.~\cite{Ajiki1994}
Some microscopic processes that contribute to the azimuthal TM mode
($\Delta n=\pm 1$) are shown in Fig.~\ref{fig:TMTE-cnt}(a,bottom) and
Fig.~\ref{fig:optical}(b). 
These intraband optical transitions are properly described by the Drude
model, and they arise only for heavy doping.}
Figure~\ref{fig:TMTE-cnt}(a,middle) shows that 
an electric dipole forms in the inner space of a CNT, which can
screen an external electric field with perpendicular polarization.
Indeed, the amplitude of the TM mode in the presence of an external electric field,
$E^{\rm ext}_{\theta,n}(r_0) e^{i(n\theta-\omega t)}$, is given by
\begin{align}
 E_{\theta,n}^{\rm TM}(r_0) = -
 \frac{E_{\theta,n}^{\rm ext}(r_0)}{1 -
 \left(\omega/\omega_n^{\rm TM}\right)^2}.
 \label{eq:ratio}
\end{align}
Thus, 
the induced and external fields have a phase difference of $\pi$ 
at low frequencies ($\omega \ll \omega_n^{\rm TM}$), and 
the total electric field (that the electrons in a CNT
actually experience), defined by 
$E_{\theta,n}(r_0)\equiv E_{\theta,n}^{\rm ext}(r_0) + E_{\theta,n}^{\rm
TM}(r_0)=E_{\theta,n}^{\rm ext}(r_0)/\left( 1 -(\omega_n^{\rm TM}/\omega
)^2\right)$,
is suppressed, as a result of the screening.
On the other hand, 
$E_{\theta,n}^{\rm TM}(r_0)$ is suppressed at high frequencies ($\omega
\gg \omega_n^{\rm TM}$) and the CNTs can absorb the polarized light.
When $\omega \simeq \omega_n^{\rm TM}$, the plasmons are resonantly
excited and form a peak structure in the absorption spectrum.

\begin{figure}[htbp]
 \begin{center}
  \includegraphics[scale=0.5]{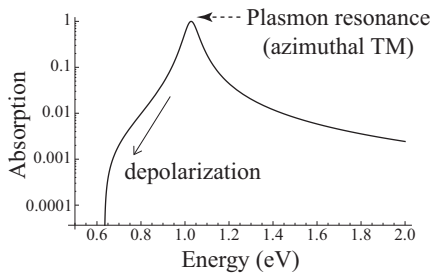}
 \end{center}
 \caption{Calculated absorption spectra as a function of light energy.
 The plot is given by Eq.~(\ref{eq:pabsorp}), 
 where $\omega_{n=1}^{\rm TM}$ is calculated from Eq.~(\ref{eq:omega_TM})
 by setting $2r_0=1.4$ nm, $\epsilon=2\epsilon_0$, $\tau_\theta=10$ fs,
 and $\sigma_\theta = 4G_0$.
 }
 \label{fig:absorp_az}
\end{figure}

Plasmon excitation by near-infrared light requires 
a heavily doped CNT 
in which more than two energy sub-bands cross the Fermi energy, 
as shown in Fig.~\ref{fig:TMTE-cnt}(a,bottom).
Because a charge-density fluctuation that causes the plasmon
excitation requires (almost) forward scatterings,~\cite{Sasaki2012b}
that is, an intraband (vertical) transition between the electronic
states with similar velocities,
the Fermi energy should be located (at least) at higher
than the bottom of the first sub-band.
Such heavily doped CNTs are already available~\cite{Kalbac2009,Hartleb2015} 
and absorption spectra
have been obtained for them.~\cite{Kazaoui1999,Igarashi2015}  
The authors of Refs.~\onlinecite{Kazaoui1999}
and~\onlinecite{Igarashi2015} observed an unidentified absorption peak
at approximately 1 eV.
We attribute it to the azimuthal TM mode, 
because Eq.~(\ref{eq:omega_TM}) satisfactorily reproduces the peak position,
$\hbar \omega_{n=\pm1}^{\rm TM}=1.027$ eV,
when we use the values reported in
Ref.~\onlinecite{Igarashi2015} ($\tau_\theta=10$ fs,~\footnote{
The electron-phonon interaction of radial breathing modes may
contribute to the lifetime $\tau_\theta$, besides the electron-electron
interaction. 
There is a strong similarity between the
electron-phonon interaction for radial breathing modes and
electron-electron interaction:
both interactions are proportional to the $2\times 2$ identity matrix in
the massless Dirac equation.~\cite{Sasaki2008c}
As a result, the generation of a charge-density fluctuation that consists
of a pair of electron and hole with different pseudospin orientations,
such as in the backward scattering, is suppressed.
To open the forward scattering channels, 
the Fermi energy should be located at higher
than the bottom of the first sub-band.
}
$\epsilon=2\epsilon_0$, and $2r_0=1.4$ nm)
with a realistic assumption $\sigma_\theta=4G_0$, 
where $G_0$ is the conductance quantum.
The factor 4 accounts for two azimuthal channels in each
valley.~\footnote{The number of effective transport channels in a
graphene is given by $|E_F|\tau/\hbar$, where $E_F$ denotes the Fermi energy.
When $|E_F|=0.65$ eV and $\tau=10$ fs, the value is approximately 10,
which is comparable to the factor 4.}
The absorption spectrum is obtained by multiplying the total field with
the dynamical conductivity as 
\begin{align}
 P(\omega) = 2 {\rm Re}\left\{
 \frac{\sigma_\theta(\omega)}{1-\left(\omega_{n=1}^{\rm TM}/\omega\right)^2} \right\}
 |E^{\rm ext}_{\theta,n=1}(r_0)|^2. 
 \label{eq:pabsorp}
\end{align}
We plot $P(\omega)$ in Fig.~\ref{fig:absorp_az}.
The calculated peak structure is consistent with the observation.~\cite{Igarashi2015}

The azimuthal plasmon has an oscillating magnetic field that is 
parallel to the tube's axis, $B_z$, as shown in the inset of Fig.~\ref{fig:TMTE-cnt}(a,middle).
Since $B_z$ is a periodic function of $\theta$ ($B_z \propto e^{in\theta}$), 
the total magnetic flux inside the cylinder vanishes at any moment in
time.
As a result, the Aharonov-Bohm effect does not
appear,~\cite{Zaric2004,Minot2004} 
and there is no notable change in the electric band structure of a CNT
due to the azimuthal plasmons.

\begin{figure}[htbp]
 \begin{center}
  \includegraphics[scale=0.5]{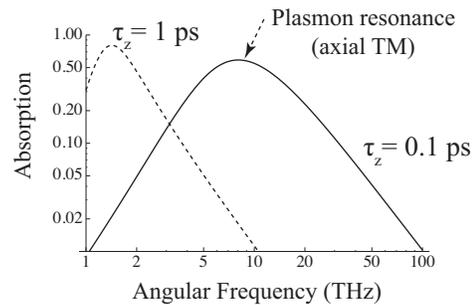}
 \end{center}
 \caption{Calculated THz absorption spectra for different $\tau_z$
 values as a function of angular frequency.
 The solid and dashed curves are for $\tau_z =0.1$ and 1 ps, respectively.~\cite{Jishi1993}
 Here, $\omega_{k}^{\rm TM}$ is calculated from Eq.~(\ref{eq:massless})
 by setting $\sigma_z = 4G_0$, $2r_0=1.4$ nm, and $\epsilon=\epsilon_0$.
 We assume a single mode with $k=\pi/L$ where $L=1$ $\mu$m is the typical length of a CNT.
 }
 \label{fig:absorp}
\end{figure}

Next, we examine the axial plasmon shown in Fig.~\ref{fig:TMTE-cnt}(b).
The axial TM mode oscillates with an angular frequency,
\begin{align}
 \omega^{\rm TM}_k = |k| \sqrt{\frac{\sigma_z r_0}{\epsilon \tau_z} \ln \left(
 \frac{2e^{-\gamma}}{|k|r_0}\right) }  - \frac{i}{2\tau_z}.
 \label{eq:massless}
\end{align}
The dispersion is basically linear in $|k|$ (apart from the small $|k|$
dependence of the logarithm) and the waves are localized,
which contrasts with the properties of the azimuthal TM mode.
The existence of this massless propagating mode can be 
understood by regarding a metallic CNT as the small waveguide.~\cite{Slepyan1999,Akima2006,Nakanishi2009,Zhang2013}
The group velocity can exceed the electron Fermi velocity 
for a metallic CNT with $2r_0=1.4$ nm, $\tau_z=0.1$ ps, and $\sigma_z=4G_0$,
and the frequencies appear in the terahertz (THz) region.
In Fig.~\ref{fig:absorp}, we show calculated THz absorption spectra that
take account of the screening effect in the axial direction.
The peak structure is consistent with several experiments.~\cite{Akima2006,Zhang2013}
Some microscopic processes in a metallic CNT that contribute to the
axial current of the TM mode are shown in Fig.~\ref{fig:TMTE-cnt}(b,bottom).

The parallel external electric field 
${\bf E}_\parallel^{\rm ext}(\omega)$
is screened by the axial TM mode.
Note, however, that since the frequency of the axial TM mode is in the THz
region, the screening is poor when $\omega$ is in the near-infrared region. 
As a result, a metallic CNT exhibits the absorption peaks of interband direct
transitions such as M$_{11}$.

Plasmons are generally sensitive to the environment.
For example, 
the plasmonic property of a CNT 
changes when a metal is present near the CNT.
Furthermore, 
since the azimuthal (axial) TM mode is an extended (localized) EM field, 
the absorption by the azimuthal mode
is more sensitive to the environment than that by the axial mode. 
This might be relevant to the fact that 
the maximum absorbance of the unidentified peak (N-band) observed in
Ref.~\onlinecite{Igarashi2015} for a doped case is 
smaller than that of the prominent peaks of the direct transitions
for an undoped case.
Contrastingly, the maximum attenuation of the broad peak in the THz
region observed in Ref.~\onlinecite{Zhang2013}
is similar to that of the prominent peaks of the
direct transitions that appear in the near-infrared region.


In summary, 
we have theoretically shown that 
a heavily doped CNT absorbs a perpendicularly polarized light in the
near-infrared region, in contrast to an undoped CNT where such an
absorption is prohibited by the screening effect caused by a
high-frequency plasmon.
The change in the optical property is because 
a heavily doped CNT can support a low-frequency plasmon (azimuthal TM mode). 
On the other hand, 
when a parallel polarization light is used to excite a metallic CNT, 
the plasmon resonance of an axial TM mode propagating along the tube
appears in the absorption spectrum in the terahertz region.
These two plasmon modes correspond to the surface plasmon of graphene
rolled up into a cylinder. 
Particularly, the azimuthal TM modes in ordinary CNTs with a diameter of
close to 1 nanometer have an advantage over graphene in terms of
wavelength down-conversion from visible light to nanoscale plasmons.

\section*{Acknowledgments}

K. S. is indebted to R. Saito and H. Kataura.

\appendix

\bibliographystyle{apsrev4-1}
%

\begin{thebibliography}{27}%
\makeatletter
\providecommand \@ifxundefined [1]{%
 \@ifx{#1\undefined}
}%
\providecommand \@ifnum [1]{%
 \ifnum #1\expandafter \@firstoftwo
 \else \expandafter \@secondoftwo
 \fi
}%
\providecommand \@ifx [1]{%
 \ifx #1\expandafter \@firstoftwo
 \else \expandafter \@secondoftwo
 \fi
}%
\providecommand \natexlab [1]{#1}%
\providecommand \enquote  [1]{``#1''}%
\providecommand \bibnamefont  [1]{#1}%
\providecommand \bibfnamefont [1]{#1}%
\providecommand \citenamefont [1]{#1}%
\providecommand \href@noop [0]{\@secondoftwo}%
\providecommand \href [0]{\begingroup \@sanitize@url \@href}%
\providecommand \@href[1]{\@@startlink{#1}\@@href}%
\providecommand \@@href[1]{\endgroup#1\@@endlink}%
\providecommand \@sanitize@url [0]{\catcode `\\12\catcode `\$12\catcode
  `\&12\catcode `\#12\catcode `\^12\catcode `\_12\catcode `\%12\relax}%
\providecommand \@@startlink[1]{}%
\providecommand \@@endlink[0]{}%
\providecommand \url  [0]{\begingroup\@sanitize@url \@url }%
\providecommand \@url [1]{\endgroup\@href {#1}{\urlprefix }}%
\providecommand \urlprefix  [0]{URL }%
\providecommand \Eprint [0]{\href }%
\@ifxundefined \urlstyle {%
  \providecommand \doi  [0]{\begingroup \@sanitize@url \@doi}%
  \providecommand \@doi [1]{\endgroup \@@startlink {\doibase
  #1}doi:\discretionary {}{}{}#1\@@endlink }%
}{%
  \providecommand \doi  [0]{doi:\discretionary{}{}{}\begingroup
  \urlstyle{rm}\Url }%
}%
\providecommand \doibase [0]{http://dx.doi.org/}%
\providecommand \Doi [0]{\begingroup \@sanitize@url \@Doi }%
\providecommand \@Doi  [1]{\endgroup\@@startlink{\doibase#1}\@@Doi}%
\providecommand \@@Doi [1]{#1\@@endlink}%
\providecommand \selectlanguage [0]{\@gobble}%
\providecommand \bibinfo  [0]{\@secondoftwo}%
\providecommand \bibfield  [0]{\@secondoftwo}%
\providecommand \translation [1]{[#1]}%
\providecommand \BibitemOpen [0]{}%
\providecommand \bibitemStop [0]{}%
\providecommand \bibitemNoStop [0]{.\EOS\space}%
\providecommand \EOS [0]{\spacefactor3000\relax}%
\providecommand \BibitemShut  [1]{\csname bibitem#1\endcsname}%
\bibitem [{\citenamefont {Iijima}(1991)}]{iijima91}%
  \BibitemOpen
  \bibfield  {author} {\bibinfo {author} {\bibfnamefont {S.}~\bibnamefont
  {Iijima}},\ }\href@noop {} {\bibfield  {journal} {\bibinfo  {journal}
  {Nature},\ }\textbf {\bibinfo {volume} {354}},\ \bibinfo {pages} {56}
  (\bibinfo {year} {1991})}\BibitemShut {NoStop}%
\bibitem [{\citenamefont {Saito}\ \emph {et~al.}(1992)\citenamefont {Saito},
  \citenamefont {Fujita}, \citenamefont {Dresselhaus},\ and\ \citenamefont
  {Dresselhaus}}]{saito92apl}%
  \BibitemOpen
  \bibfield  {author} {\bibinfo {author} {\bibfnamefont {R.}~\bibnamefont
  {Saito}}, \bibinfo {author} {\bibfnamefont {M.}~\bibnamefont {Fujita}},
  \bibinfo {author} {\bibfnamefont {G.}~\bibnamefont {Dresselhaus}}, \ and\
  \bibinfo {author} {\bibfnamefont {M.~S.}\ \bibnamefont {Dresselhaus}},\
  }\href@noop {} {\bibfield  {journal} {\bibinfo  {journal} {Appl. Phys.
  Lett.},\ }\textbf {\bibinfo {volume} {60}},\ \bibinfo {pages} {2204}
  (\bibinfo {year} {1992})}\BibitemShut {NoStop}%
\bibitem [{\citenamefont {Ichida}\ \emph {et~al.}(2004)\citenamefont {Ichida},
  \citenamefont {Mizuno}, \citenamefont {Kataura}, \citenamefont {Achiba},\
  and\ \citenamefont {Nakamura}}]{ichida04}%
  \BibitemOpen
  \bibfield  {author} {\bibinfo {author} {\bibfnamefont {M.}~\bibnamefont
  {Ichida}}, \bibinfo {author} {\bibfnamefont {S.}~\bibnamefont {Mizuno}},
  \bibinfo {author} {\bibfnamefont {H.}~\bibnamefont {Kataura}}, \bibinfo
  {author} {\bibfnamefont {Y.}~\bibnamefont {Achiba}}, \ and\ \bibinfo {author}
  {\bibfnamefont {A.}~\bibnamefont {Nakamura}},\ }\href@noop {} {\bibfield
  {journal} {\bibinfo  {journal} {Applied Physics A},\ }\textbf {\bibinfo
  {volume} {78}},\ \bibinfo {pages} {1117} (\bibinfo {year}
  {2004})}\BibitemShut {NoStop}%
\bibitem [{\citenamefont {Dresselhaus}\ \emph {et~al.}(2010)\citenamefont
  {Dresselhaus}, \citenamefont {Jorio}, \citenamefont {Hofmann}, \citenamefont
  {Dresselhaus},\ and\ \citenamefont {Saito}}]{Dresselhaus2010}%
  \BibitemOpen
  \bibfield  {author} {\bibinfo {author} {\bibfnamefont {M.~S.}\ \bibnamefont
  {Dresselhaus}}, \bibinfo {author} {\bibfnamefont {A.}~\bibnamefont {Jorio}},
  \bibinfo {author} {\bibfnamefont {M.}~\bibnamefont {Hofmann}}, \bibinfo
  {author} {\bibfnamefont {G.}~\bibnamefont {Dresselhaus}}, \ and\ \bibinfo
  {author} {\bibfnamefont {R.}~\bibnamefont {Saito}},\ }\Doi
  {10.1021/nl904286r} {\bibfield  {journal} {\bibinfo  {journal} {Nano
  letters},\ }\textbf {\bibinfo {volume} {10}},\ \bibinfo {pages} {751}
  (\bibinfo {year} {2010})},\ ISSN \bibinfo {issn} {1530-6992}\BibitemShut
  {NoStop}%
\bibitem [{\citenamefont {Stern}(1967)}]{stern67}%
  \BibitemOpen
  \bibfield  {author} {\bibinfo {author} {\bibfnamefont {F.}~\bibnamefont
  {Stern}},\ }\Doi {10.1103/PhysRevLett.18.546} {\bibfield  {journal} {\bibinfo
   {journal} {Physical Review Letters},\ }\textbf {\bibinfo {volume} {18}},\
  \bibinfo {pages} {546} (\bibinfo {year} {1967})},\ ISSN \bibinfo {issn}
  {0031-9007}\BibitemShut {NoStop}%
\bibitem [{\citenamefont {Wunsch}\ \emph {et~al.}(2006)\citenamefont {Wunsch},
  \citenamefont {Stauber}, \citenamefont {Sols},\ and\ \citenamefont
  {Guinea}}]{Wunsch2006}%
  \BibitemOpen
  \bibfield  {author} {\bibinfo {author} {\bibfnamefont {B.}~\bibnamefont
  {Wunsch}}, \bibinfo {author} {\bibfnamefont {T.}~\bibnamefont {Stauber}},
  \bibinfo {author} {\bibfnamefont {F.}~\bibnamefont {Sols}}, \ and\ \bibinfo
  {author} {\bibfnamefont {F.}~\bibnamefont {Guinea}},\ }\Doi
  {10.1088/1367-2630/8/12/318} {\bibfield  {journal} {\bibinfo  {journal} {New
  Journal of Physics},\ }\textbf {\bibinfo {volume} {8}},\ \bibinfo {pages}
  {318} (\bibinfo {year} {2006})},\ ISSN \bibinfo {issn}
  {1367-2630}\BibitemShut {NoStop}%
\bibitem [{\citenamefont {Hwang}\ and\ \citenamefont {{Das
  Sarma}}(2007)}]{hwang07}%
  \BibitemOpen
  \bibfield  {author} {\bibinfo {author} {\bibfnamefont {E.~H.}\ \bibnamefont
  {Hwang}}\ and\ \bibinfo {author} {\bibfnamefont {S.}~\bibnamefont {{Das
  Sarma}}},\ }\Doi {10.1103/PhysRevB.75.205418} {\bibfield  {journal} {\bibinfo
   {journal} {Physical Review B},\ }\textbf {\bibinfo {volume} {75}},\ \bibinfo
  {pages} {205418} (\bibinfo {year} {2007})},\ ISSN \bibinfo {issn}
  {1098-0121}\BibitemShut {NoStop}%
\bibitem [{\citenamefont {Jablan}\ \emph {et~al.}(2009)\citenamefont {Jablan},
  \citenamefont {Buljan},\ and\ \citenamefont
  {Solja{\v{c}}i{\'{c}}}}]{Jablan2009}%
  \BibitemOpen
  \bibfield  {author} {\bibinfo {author} {\bibfnamefont {M.}~\bibnamefont
  {Jablan}}, \bibinfo {author} {\bibfnamefont {H.}~\bibnamefont {Buljan}}, \
  and\ \bibinfo {author} {\bibfnamefont {M.}~\bibnamefont
  {Solja{\v{c}}i{\'{c}}}},\ }\Doi {10.1103/PhysRevB.80.245435} {\bibfield
  {journal} {\bibinfo  {journal} {Physical Review B},\ }\textbf {\bibinfo
  {volume} {80}},\ \bibinfo {pages} {245435} (\bibinfo {year} {2009})},\ ISSN
  \bibinfo {issn} {1098-0121}\BibitemShut {NoStop}%
\bibitem [{\citenamefont {Ajiki}\ and\ \citenamefont {Ando}(1994)}]{Ajiki1994}%
  \BibitemOpen
  \bibfield  {author} {\bibinfo {author} {\bibfnamefont {H.}~\bibnamefont
  {Ajiki}}\ and\ \bibinfo {author} {\bibfnamefont {T.}~\bibnamefont {Ando}},\
  }\Doi {10.1016/0921-4526(94)91112-6} {\bibfield  {journal} {\bibinfo
  {journal} {Physica B: Condensed Matter},\ }\textbf {\bibinfo {volume}
  {201}},\ \bibinfo {pages} {349} (\bibinfo {year} {1994})},\ ISSN \bibinfo
  {issn} {09214526}\BibitemShut {NoStop}%
\bibitem [{\citenamefont {Sasaki}\ \emph {et~al.}(2011)\citenamefont {Sasaki},
  \citenamefont {Kato}, \citenamefont {Tokura}, \citenamefont {Oguri},\ and\
  \citenamefont {Sogawa}}]{Sasaki2011}%
  \BibitemOpen
  \bibfield  {author} {\bibinfo {author} {\bibfnamefont {K.-i.}\ \bibnamefont
  {Sasaki}}, \bibinfo {author} {\bibfnamefont {K.}~\bibnamefont {Kato}},
  \bibinfo {author} {\bibfnamefont {Y.}~\bibnamefont {Tokura}}, \bibinfo
  {author} {\bibfnamefont {K.}~\bibnamefont {Oguri}}, \ and\ \bibinfo {author}
  {\bibfnamefont {T.}~\bibnamefont {Sogawa}},\ }\Doi
  {10.1103/PhysRevB.84.085458} {\bibfield  {journal} {\bibinfo  {journal}
  {Physical Review B},\ }\textbf {\bibinfo {volume} {84}},\ \bibinfo {pages}
  {085458} (\bibinfo {year} {2011})},\ ISSN \bibinfo {issn}
  {1098-0121}\BibitemShut {NoStop}%
\bibitem [{\citenamefont {St{\'{e}}phan}\ \emph {et~al.}(2002)\citenamefont
  {St{\'{e}}phan}, \citenamefont {Taverna}, \citenamefont {Kociak},
  \citenamefont {Suenaga}, \citenamefont {Henrard},\ and\ \citenamefont
  {Colliex}}]{Stephan2002}%
  \BibitemOpen
  \bibfield  {author} {\bibinfo {author} {\bibfnamefont {O.}~\bibnamefont
  {St{\'{e}}phan}}, \bibinfo {author} {\bibfnamefont {D.}~\bibnamefont
  {Taverna}}, \bibinfo {author} {\bibfnamefont {M.}~\bibnamefont {Kociak}},
  \bibinfo {author} {\bibfnamefont {K.}~\bibnamefont {Suenaga}}, \bibinfo
  {author} {\bibfnamefont {L.}~\bibnamefont {Henrard}}, \ and\ \bibinfo
  {author} {\bibfnamefont {C.}~\bibnamefont {Colliex}},\ }\Doi
  {10.1103/PhysRevB.66.155422} {\bibfield  {journal} {\bibinfo  {journal}
  {Physical Review B},\ }\textbf {\bibinfo {volume} {66}},\ \bibinfo {pages}
  {155422} (\bibinfo {year} {2002})},\ ISSN \bibinfo {issn}
  {0163-1829}\BibitemShut {NoStop}%
\bibitem [{\citenamefont {{Garc{\'{i}}a de Abajo}}(2010)}]{GarciadeAbajo2010}%
  \BibitemOpen
  \bibfield  {author} {\bibinfo {author} {\bibfnamefont {F.~J.}\ \bibnamefont
  {{Garc{\'{i}}a de Abajo}}},\ }\Doi {10.1103/RevModPhys.82.209} {\bibfield
  {journal} {\bibinfo  {journal} {Reviews of Modern Physics},\ }\textbf
  {\bibinfo {volume} {82}},\ \bibinfo {pages} {209} (\bibinfo {year} {2010})},\
  ISSN \bibinfo {issn} {0034-6861}\BibitemShut {NoStop}%
\bibitem [{\citenamefont {Nakayama}(1974)}]{nakayama74}%
  \BibitemOpen
  \bibfield  {author} {\bibinfo {author} {\bibfnamefont {M.}~\bibnamefont
  {Nakayama}},\ }\Doi {10.1143/JPSJ.36.393} {\bibfield  {journal} {\bibinfo
  {journal} {J. Phys. Soc. Jpn.},\ }\textbf {\bibinfo {volume} {36}},\ \bibinfo
  {pages} {393} (\bibinfo {year} {1974})}\BibitemShut {NoStop}%
\bibitem [{\citenamefont {Sasaki}\ \emph {et~al.}(2016)\citenamefont {Sasaki},
  \citenamefont {Murakami}, \citenamefont {Tokura},\ and\ \citenamefont
  {Yamamoto}}]{Sasaki2016}%
  \BibitemOpen
  \bibfield  {author} {\bibinfo {author} {\bibfnamefont {K.-i.}\ \bibnamefont
  {Sasaki}}, \bibinfo {author} {\bibfnamefont {S.}~\bibnamefont {Murakami}},
  \bibinfo {author} {\bibfnamefont {Y.}~\bibnamefont {Tokura}}, \ and\ \bibinfo
  {author} {\bibfnamefont {H.}~\bibnamefont {Yamamoto}},\ }\Doi
  {10.1103/PhysRevB.93.125402} {\bibfield  {journal} {\bibinfo  {journal}
  {Physical Review B},\ }\textbf {\bibinfo {volume} {93}},\ \bibinfo {pages}
  {125402} (\bibinfo {year} {2016})},\ ISSN \bibinfo {issn}
  {2469-9950}\BibitemShut {NoStop}%
\bibitem [{\citenamefont {Sasaki}\ \emph {et~al.}(2012)\citenamefont {Sasaki},
  \citenamefont {Kato}, \citenamefont {Tokura}, \citenamefont {Suzuki},\ and\
  \citenamefont {Sogawa}}]{Sasaki2012b}%
  \BibitemOpen
  \bibfield  {author} {\bibinfo {author} {\bibfnamefont {K.-i.}\ \bibnamefont
  {Sasaki}}, \bibinfo {author} {\bibfnamefont {K.}~\bibnamefont {Kato}},
  \bibinfo {author} {\bibfnamefont {Y.}~\bibnamefont {Tokura}}, \bibinfo
  {author} {\bibfnamefont {S.}~\bibnamefont {Suzuki}}, \ and\ \bibinfo {author}
  {\bibfnamefont {T.}~\bibnamefont {Sogawa}},\ }\Doi
  {10.1103/PhysRevB.86.201403} {\bibfield  {journal} {\bibinfo  {journal}
  {Physical Review B},\ }\textbf {\bibinfo {volume} {86}},\ \bibinfo {pages}
  {201403} (\bibinfo {year} {2012})},\ ISSN \bibinfo {issn}
  {1098-0121}\BibitemShut {NoStop}%
\bibitem [{\citenamefont {Kalbac}\ \emph {et~al.}(2009)\citenamefont {Kalbac},
  \citenamefont {Farhat}, \citenamefont {Kavan}, \citenamefont {Kong},
  \citenamefont {Sasaki}, \citenamefont {Saito},\ and\ \citenamefont
  {Dresselhaus}}]{Kalbac2009}%
  \BibitemOpen
  \bibfield  {author} {\bibinfo {author} {\bibfnamefont {M.}~\bibnamefont
  {Kalbac}}, \bibinfo {author} {\bibfnamefont {H.}~\bibnamefont {Farhat}},
  \bibinfo {author} {\bibfnamefont {L.}~\bibnamefont {Kavan}}, \bibinfo
  {author} {\bibfnamefont {J.}~\bibnamefont {Kong}}, \bibinfo {author}
  {\bibfnamefont {K.-i.}\ \bibnamefont {Sasaki}}, \bibinfo {author}
  {\bibfnamefont {R.}~\bibnamefont {Saito}}, \ and\ \bibinfo {author}
  {\bibfnamefont {M.~S.}\ \bibnamefont {Dresselhaus}},\ }\Doi
  {10.1021/nn9004318} {\bibfield  {journal} {\bibinfo  {journal} {ACS Nano},\
  }\textbf {\bibinfo {volume} {3}},\ \bibinfo {pages} {2320} (\bibinfo {year}
  {2009})},\ ISSN \bibinfo {issn} {1936-0851}\BibitemShut {NoStop}%
\bibitem [{\citenamefont {Hartleb}\ \emph {et~al.}(2015)\citenamefont
  {Hartleb}, \citenamefont {Sp{\"{a}}th},\ and\ \citenamefont
  {Hertel}}]{Hartleb2015}%
  \BibitemOpen
  \bibfield  {author} {\bibinfo {author} {\bibfnamefont {H.}~\bibnamefont
  {Hartleb}}, \bibinfo {author} {\bibfnamefont {F.}~\bibnamefont
  {Sp{\"{a}}th}}, \ and\ \bibinfo {author} {\bibfnamefont {T.}~\bibnamefont
  {Hertel}},\ }\Doi {10.1021/acsnano.5b04707} {\bibfield  {journal} {\bibinfo
  {journal} {ACS nano},\ }\textbf {\bibinfo {volume} {9}},\ \bibinfo {pages}
  {10461} (\bibinfo {year} {2015})},\ ISSN \bibinfo {issn}
  {1936-086X}\BibitemShut {NoStop}%
\bibitem [{\citenamefont {Kazaoui}\ \emph {et~al.}(1999)\citenamefont
  {Kazaoui}, \citenamefont {Minami}, \citenamefont {Jacquemin}, \citenamefont
  {Kataura},\ and\ \citenamefont {Achiba}}]{Kazaoui1999}%
  \BibitemOpen
  \bibfield  {author} {\bibinfo {author} {\bibfnamefont {S.}~\bibnamefont
  {Kazaoui}}, \bibinfo {author} {\bibfnamefont {N.}~\bibnamefont {Minami}},
  \bibinfo {author} {\bibfnamefont {R.}~\bibnamefont {Jacquemin}}, \bibinfo
  {author} {\bibfnamefont {H.}~\bibnamefont {Kataura}}, \ and\ \bibinfo
  {author} {\bibfnamefont {Y.}~\bibnamefont {Achiba}},\ }\Doi
  {10.1103/PhysRevB.60.13339} {\bibfield  {journal} {\bibinfo  {journal}
  {Physical Review B},\ }\textbf {\bibinfo {volume} {60}},\ \bibinfo {pages}
  {13339} (\bibinfo {year} {1999})},\ ISSN \bibinfo {issn}
  {0163-1829}\BibitemShut {NoStop}%
\bibitem [{\citenamefont {Igarashi}\ \emph {et~al.}(2015)\citenamefont
  {Igarashi}, \citenamefont {Kawai}, \citenamefont {Yanagi}, \citenamefont
  {Cuong}, \citenamefont {Okada},\ and\ \citenamefont
  {Pichler}}]{Igarashi2015}%
  \BibitemOpen
  \bibfield  {author} {\bibinfo {author} {\bibfnamefont {T.}~\bibnamefont
  {Igarashi}}, \bibinfo {author} {\bibfnamefont {H.}~\bibnamefont {Kawai}},
  \bibinfo {author} {\bibfnamefont {K.}~\bibnamefont {Yanagi}}, \bibinfo
  {author} {\bibfnamefont {N.~T.}\ \bibnamefont {Cuong}}, \bibinfo {author}
  {\bibfnamefont {S.}~\bibnamefont {Okada}}, \ and\ \bibinfo {author}
  {\bibfnamefont {T.}~\bibnamefont {Pichler}},\ }\Doi
  {10.1103/PhysRevLett.114.176807} {\bibfield  {journal} {\bibinfo  {journal}
  {Physical Review Letters},\ }\textbf {\bibinfo {volume} {114}},\ \bibinfo
  {pages} {176807} (\bibinfo {year} {2015})},\ ISSN \bibinfo {issn}
  {0031-9007}\BibitemShut {NoStop}%
\bibitem [{\citenamefont {Zaric}(2004)}]{Zaric2004}%
  \BibitemOpen
  \bibfield  {author} {\bibinfo {author} {\bibfnamefont {S.}~\bibnamefont
  {Zaric}},\ }\Doi {10.1126/science.1096524} {\bibfield  {journal} {\bibinfo
  {journal} {Science},\ }\textbf {\bibinfo {volume} {304}},\ \bibinfo {pages}
  {1129} (\bibinfo {year} {2004})},\ ISSN \bibinfo {issn}
  {0036-8075}\BibitemShut {NoStop}%
\bibitem [{\citenamefont {Minot}\ \emph {et~al.}(2004)\citenamefont {Minot},
  \citenamefont {Yaish}, \citenamefont {Sazonova},\ and\ \citenamefont
  {McEuen}}]{Minot2004}%
  \BibitemOpen
  \bibfield  {author} {\bibinfo {author} {\bibfnamefont {E.~D.}\ \bibnamefont
  {Minot}}, \bibinfo {author} {\bibfnamefont {Y.}~\bibnamefont {Yaish}},
  \bibinfo {author} {\bibfnamefont {V.}~\bibnamefont {Sazonova}}, \ and\
  \bibinfo {author} {\bibfnamefont {P.~L.}\ \bibnamefont {McEuen}},\ }\Doi
  {10.1038/nature02425} {\bibfield  {journal} {\bibinfo  {journal} {Nature},\
  }\textbf {\bibinfo {volume} {428}},\ \bibinfo {pages} {536} (\bibinfo {year}
  {2004})},\ ISSN \bibinfo {issn} {1476-4687}\BibitemShut {NoStop}%
\bibitem [{\citenamefont {Slepyan}\ \emph {et~al.}(1999)\citenamefont
  {Slepyan}, \citenamefont {Maksimenko}, \citenamefont {Lakhtakia},
  \citenamefont {Yevtushenko},\ and\ \citenamefont {Gusakov}}]{Slepyan1999}%
  \BibitemOpen
  \bibfield  {author} {\bibinfo {author} {\bibfnamefont {G.~Y.}\ \bibnamefont
  {Slepyan}}, \bibinfo {author} {\bibfnamefont {S.~A.}\ \bibnamefont
  {Maksimenko}}, \bibinfo {author} {\bibfnamefont {A.}~\bibnamefont
  {Lakhtakia}}, \bibinfo {author} {\bibfnamefont {O.}~\bibnamefont
  {Yevtushenko}}, \ and\ \bibinfo {author} {\bibfnamefont {A.~V.}\ \bibnamefont
  {Gusakov}},\ }\Doi {10.1103/PhysRevB.60.17136} {\bibfield  {journal}
  {\bibinfo  {journal} {Physical Review B},\ }\textbf {\bibinfo {volume}
  {60}},\ \bibinfo {pages} {17136} (\bibinfo {year} {1999})},\ ISSN \bibinfo
  {issn} {0163-1829}\BibitemShut {NoStop}%
\bibitem [{\citenamefont {Akima}\ \emph {et~al.}(2006)\citenamefont {Akima},
  \citenamefont {Iwasa}, \citenamefont {Brown}, \citenamefont {Barbour},
  \citenamefont {Cao}, \citenamefont {Musfeldt}, \citenamefont {Matsui},
  \citenamefont {Toyota}, \citenamefont {Shiraishi}, \citenamefont {Shimoda},\
  and\ \citenamefont {Zhou}}]{Akima2006}%
  \BibitemOpen
  \bibfield  {author} {\bibinfo {author} {\bibfnamefont {N.}~\bibnamefont
  {Akima}}, \bibinfo {author} {\bibfnamefont {Y.}~\bibnamefont {Iwasa}},
  \bibinfo {author} {\bibfnamefont {S.}~\bibnamefont {Brown}}, \bibinfo
  {author} {\bibfnamefont {A.~M.}\ \bibnamefont {Barbour}}, \bibinfo {author}
  {\bibfnamefont {J.}~\bibnamefont {Cao}}, \bibinfo {author} {\bibfnamefont
  {J.~L.}\ \bibnamefont {Musfeldt}}, \bibinfo {author} {\bibfnamefont
  {H.}~\bibnamefont {Matsui}}, \bibinfo {author} {\bibfnamefont
  {N.}~\bibnamefont {Toyota}}, \bibinfo {author} {\bibfnamefont
  {M.}~\bibnamefont {Shiraishi}}, \bibinfo {author} {\bibfnamefont
  {H.}~\bibnamefont {Shimoda}}, \ and\ \bibinfo {author} {\bibfnamefont
  {O.}~\bibnamefont {Zhou}},\ }\Doi {10.1002/adma.200502505} {\bibfield
  {journal} {\bibinfo  {journal} {Advanced Materials},\ }\textbf {\bibinfo
  {volume} {18}},\ \bibinfo {pages} {1166} (\bibinfo {year} {2006})},\ ISSN
  \bibinfo {issn} {0935-9648}\BibitemShut {NoStop}%
\bibitem [{\citenamefont {Nakanishi}\ and\ \citenamefont
  {Ando}(2009)}]{Nakanishi2009}%
  \BibitemOpen
  \bibfield  {author} {\bibinfo {author} {\bibfnamefont {T.}~\bibnamefont
  {Nakanishi}}\ and\ \bibinfo {author} {\bibfnamefont {T.}~\bibnamefont
  {Ando}},\ }\Doi {10.1143/JPSJ.78.114708} {\bibfield  {journal} {\bibinfo
  {journal} {Journal of the Physical Society of Japan},\ }\textbf {\bibinfo
  {volume} {78}},\ \bibinfo {pages} {114708} (\bibinfo {year} {2009})},\ ISSN
  \bibinfo {issn} {0031-9015}\BibitemShut {NoStop}%
\bibitem [{\citenamefont {Zhang}\ \emph {et~al.}(2013)\citenamefont {Zhang},
  \citenamefont {H{\'{a}}roz}, \citenamefont {Jin}, \citenamefont {Ren},
  \citenamefont {Wang}, \citenamefont {Arvidson}, \citenamefont
  {L{\"{u}}ttge},\ and\ \citenamefont {Kono}}]{Zhang2013}%
  \BibitemOpen
  \bibfield  {author} {\bibinfo {author} {\bibfnamefont {Q.}~\bibnamefont
  {Zhang}}, \bibinfo {author} {\bibfnamefont {E.~H.}\ \bibnamefont
  {H{\'{a}}roz}}, \bibinfo {author} {\bibfnamefont {Z.}~\bibnamefont {Jin}},
  \bibinfo {author} {\bibfnamefont {L.}~\bibnamefont {Ren}}, \bibinfo {author}
  {\bibfnamefont {X.}~\bibnamefont {Wang}}, \bibinfo {author} {\bibfnamefont
  {R.~S.}\ \bibnamefont {Arvidson}}, \bibinfo {author} {\bibfnamefont
  {A.}~\bibnamefont {L{\"{u}}ttge}}, \ and\ \bibinfo {author} {\bibfnamefont
  {J.}~\bibnamefont {Kono}},\ }\Doi {10.1021/nl403175g} {\bibfield  {journal}
  {\bibinfo  {journal} {Nano letters},\ }\textbf {\bibinfo {volume} {13}},\
  \bibinfo {pages} {5991} (\bibinfo {year} {2013})},\ ISSN \bibinfo {issn}
  {1530-6992}\BibitemShut {NoStop}%
\bibitem [{\citenamefont {Jishi}\ \emph {et~al.}(1993)\citenamefont {Jishi},
  \citenamefont {Dresselhaus},\ and\ \citenamefont {Dresselhaus}}]{Jishi1993}%
  \BibitemOpen
  \bibfield  {author} {\bibinfo {author} {\bibfnamefont {R.~A.}\ \bibnamefont
  {Jishi}}, \bibinfo {author} {\bibfnamefont {M.~S.}\ \bibnamefont
  {Dresselhaus}}, \ and\ \bibinfo {author} {\bibfnamefont {G.}~\bibnamefont
  {Dresselhaus}},\ }\Doi {10.1103/PhysRevB.48.11385} {\bibfield  {journal}
  {\bibinfo  {journal} {Physical Review B},\ }\textbf {\bibinfo {volume}
  {48}},\ \bibinfo {pages} {11385} (\bibinfo {year} {1993})},\ ISSN \bibinfo
  {issn} {0163-1829}\BibitemShut {NoStop}%
\bibitem [{\citenamefont {Sasaki}\ \emph {et~al.}(2008)\citenamefont {Sasaki},
  \citenamefont {Saito}, \citenamefont {Dresselhaus}, \citenamefont
  {Dresselhaus}, \citenamefont {Farhat},\ and\ \citenamefont
  {Kong}}]{Sasaki2008c}%
  \BibitemOpen
  \bibfield  {author} {\bibinfo {author} {\bibfnamefont {K.-i.}\ \bibnamefont
  {Sasaki}}, \bibinfo {author} {\bibfnamefont {R.}~\bibnamefont {Saito}},
  \bibinfo {author} {\bibfnamefont {G.}~\bibnamefont {Dresselhaus}}, \bibinfo
  {author} {\bibfnamefont {M.~S.}\ \bibnamefont {Dresselhaus}}, \bibinfo
  {author} {\bibfnamefont {H.}~\bibnamefont {Farhat}}, \ and\ \bibinfo {author}
  {\bibfnamefont {J.}~\bibnamefont {Kong}},\ }\Doi {10.1103/PhysRevB.78.235405}
  {\bibfield  {journal} {\bibinfo  {journal} {Physical Review B},\ }\textbf
  {\bibinfo {volume} {78}},\ \bibinfo {pages} {235405} (\bibinfo {year}
  {2008})},\ ISSN \bibinfo {issn} {1098-0121}\BibitemShut {NoStop}%
\end{thebibliography}

%

\end{document}